\newcommand{\PBVO}{Pb$_2$VO(PO$_4$)$_2$\xspace}

\newcommand{\be}{\begin{equation} }
\newcommand{\ee}{\end{equation} }
\newcommand{\bea}{\begin{eqnarray} }
\newcommand{\eea}{\end{eqnarray} }

\newcommand{\PbV}{Pb$_2$VO(PO$_4$)$_2$\xspace}

\documentclass[aps,prb,nofootinbib,twocolumn,superscriptaddress]{revtex4-1}

\usepackage{amsmath,amssymb,bm}
\usepackage{graphicx}
\usepackage{epspdfconversion}
\usepackage{color}
\usepackage{xspace}
\usepackage[utf8]{inputenc}

\begin{document}

%\preprint{}

\title{Magnetic structure and spin waves in the frustrated ferro-antiferromagnet \PBVO}

\author{S. Bettler}
\email{sbettler@phys.ethz.ch}
\author{F. Landolt}
\author{\"O. M. Aksoy}
\author{Z. Yan}
\author{S. Gvasaliya}
\affiliation{Laboratory for Solid State Physics, ETH Z\"urich, 8093 Z\"urich, Switzerland.}
\author{Y. Qiu}
\affiliation{NIST Centre for Neutron Research, National Institute of Standards and Technology, Gaithersburg, Maryland 20878, USA.}
\author{E. Ressouche}
\author{K. Beauvois}
\author{S. Raymond}
\affiliation{Univ. Grenoble Alpes, CEA, INAC-MEM, 38000 Grenoble, France.}
\author{A. N. Ponomaryov}
\author{S. A. Zvyagin}
\affiliation{Dresden High Magnetic Field Laboratory (HLD-EMFL), Helmholtz-Zentrum Dresden-Rossendorf, 01328 Dresden, Germany}
\author{A. Zheludev}
\email{zhelud@ethz.ch}
\homepage{http://www.neutron.ethz.ch/}
\affiliation{Laboratory for Solid State Physics, ETH Z\"urich, 8093 Z\"urich, Switzerland.}

\date{\today}

\begin{abstract}
Single crystal neutron diffraction, inelastic neutron scattering and electron spin resonance experiments are used to study the magnetic structure and spin waves in  \PbV, a prototypical layered $S=1/2$ ferromagnet with frustrating next nearest neighbor antiferromagnetic interactions. The observed excitation spectrum is found to be inconsistent with a simple square lattice model previously proposed for this material. At least four distinct exchange coupling constants are required to reproduce the measured spin wave dispersion. The degree of magnetic frustration is correspondingly revised and found to be substantially smaller than in all previous estimates.
\end{abstract}

%\keywords{}

\maketitle
\section{Introduction}
In classical spin systems, geometric frustration is usually resolved by the ``least frustrated'' magnetically ordered state. The actual ground state spin configuration depends on the type and 
strength of frustration. As a function of this parameter, one  expects transitions between
different ordered phases. These ``classical'' transition points  are a
promising area to look for novel {\it quantum} phases and
excitations.\cite{Starykh2015} A case in point is the 
Heisenberg ferromagnet on a square lattice frustrated
by antiferromagnetic (AF) next-nearest neighbor coupling. The classical model is either
a collinear ferromagnet or a   ``columnar'' AF (CAF) structure.  The $S=1/2$ quantum model near the transition between these classical states, for $-0.7<J_2/J_1<-0.4$, is predicted to be a so-called spin nematic.\cite{Shannon2006,Ueda2007,Shindou2013} This exotic quantum phase shows no conventional dipolar magnetic order but features spontaneously anisotropic quantum spin fluctuations.  In strong enough applied
fields a nematic phase is supported for all $J_2/J_1<- 0.4$.\cite{Ueda2013,Ueda2015}

To this day, the only known potential experimental realizations of the  ferro-antiferromagnet
square lattice model are among layered 
vanadophosphates.\cite{Tsirlin2009} Our understanding of these compounds is far from complete though, since most experiments to date were done on powder samples.
The only neutron work reported to date is on \PbV, which is perhaps the most thoroughly studied member of the series.\cite{Skoulatos2007,Skoulatos2009}
These powder neutron diffraction experiments confirmed a magnetically ordered CAF structure. The exchange constants $J_1= -0.52$~meV and $J_2=0.84$~meV were first estimated from magnetic susceptibility data.\cite{Kaul2004} These values are reasonably consistent with the frustration ratio deduced from  quasielastic diffuse magnetic neutron scattering assuming a perfect square lattice model:\cite{Skoulatos2009} $J_2/J_1\sim-2.4$. \PbV was also probed by  Q-band electron spin resonance (ESR)~\cite{Forster2013}, NMR~\cite{Nath2009} and magnetometry as well as specific heat~\cite{Kaul2005} experiments on small single crystal samples. The consensus was that this compound is a good ``baseline'' ferro-antiferromagnet square lattice system, substantially frustrated but still in the classical CAF phase.

By their nature, previous bulk measurements and powder neutron diffraction experiments were unable to assess just how well \PbV corresponds to the ferro-antiferromagnet square lattice model in the first place. This is a valid point of concern, since the material is monoclinic rather than tetragonal. Therefore, the spins no longer form a perfect square. The spin arrangement is not even rectangular. There are as many as four atoms per crystallographic unit cell with several inequivalent bonds between them. In the present work we report the results of high-field ESR, neutron diffraction and inelastic neutron scattering measurements on {\em single crystal} samples of this compound. We show that excitations deviate significantly from those in a square lattice columnar state and require at least four distinct exchange constants to describe them. In addition, we accurately determine the direction and magnitude of the ordered moment at low temperatures and confirm the exceptional magnetic two-dimensionality of this material.
We conclude that it indeed features ferro-antiferromagnet frustration, but deviates substantially from a simplistic square-lattice description.

\section{Experimental}
\label{structure}
\begin{figure}
\includegraphics[width=\columnwidth]{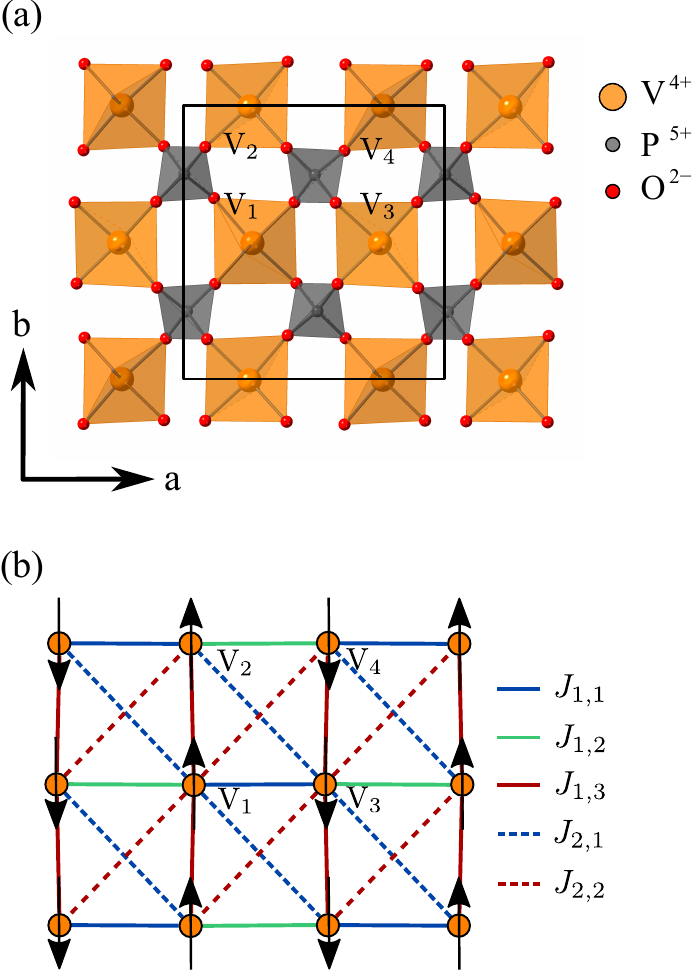}
\caption{(a) A single vanadophosphate layer in the crystal structure of \PbV showing the crystallographic unit cell and the four distinct $V^{4+}$ ions that it contains.  (b) The five distinct nearest-neighbor and next-nearest-neighbor V-V bonds in each such layer (lines). Arrows represent the spin orientation in the ordered state as deduced from our single crystal diffraction data.}
\label{crystal}
\end{figure}

\PBVO crystallizes in space group P12$_1$/a1 (No.14) with lattice parameters a=8.747(4), b=9.016(5), c=9.863(9) \AA , $\beta$=100.96(4)$^{\circ}$.\cite{Shpanchenko2006} Each unit cell contains four formula units. 
The crystal structure features layers of VO$_5$ pyramids connected through PO$_4$ tetrahedra in the $(ab)$ plane [Fig.~\ref{crystal}(a)]. The V-layers are separated by phosphate tetrahedra and lead ions. Within each layer, symmetry allows for three distinct nearest neighbor (denoted $J_{1,1}$, $J_{1,2}$ and $J_{1,3}$, respectively) and two next-nearest neighbor ($J_{2,1}$ and $J_{2,2}$) interactions, as shown in Fig. \ref{crystal}(b). The exchange paths for nearest neighbor ferromagnetic interactions as well as next-nearest neighbor antiferromagnetic exchanges both run through two bridging oxygen atoms. The individual V-O-O bridging angles along the ferromagnetic and antiferromagnetic exchange paths are in the ranges of 100 to 120 degrees and 160 to 180 degrees respectively.\cite{Shpanchenko2006}

Powders of Pb$_2$(VO)(PO$_4$)$_2$ were synthesized in an Ar-flow muffle furnace using stoichiometric amounts of high-purity commercially available precursors of PbO(99.999\%), NH$_4$H$_2$PO$_4$(99.999\%), V$_2$O$_3$(99.99\%) and V$_2$O$_5$(99.99\%)(Sigma-Aldrich\footnote{The identification of any commercial product or trade name does not imply endorsement or recommendation by the National Institute of Standards and Technology.}). Single crystals were grown in a quartz tube using the self-flux Bridgman method in a modified vertical muffle furnace. During the growing period, care was taken to prevent ingress of oxygen.
For neutron diffraction measurements we employed a 23~mg single crystal sample with mosaic spread of 0.6$^\circ$ full width at half maximum. Inelastic neutron scattering was done on a 1.5~g single crystal sample with 0.6$^\circ$ mosaic spread. For ESR experiments single-crystalline samples of Pb$_2$(VO)(PO$_4$)$_2$ with typical sizes of 3x3x1 mm$^3$ were used.

The magnetic structure was determined on the CEA-CRG D23 diffractometer at ILL using neutrons of a wavelength $\lambda=2.35$~\AA\xspace. Sample environment was a standard orange cryostat. A pyrolytic graphite filter was used to suppress higher harmonics.
Magnetic ordering corresponds to a $(0,0,0)$ propagation vector (recall that there are  4 magnetic atoms per unit cell). As a result, many magnetic Bragg reflections coincide with nuclear peaks. Magnetic reflections for which this was not the case were measured in usual rocking scans at 1.5~K counting about 45~s/point. For the remaining magnetic reflections we first measured the rocking curve of the underlying nuclear peak at $T=5~\mathrm{K}>T_N\sim 3.5$~K. Each scan was then analyzed using a Gaussian fit. Subsequently, the combined nuclear and magnetic contributions were measured at $T=1.5$~K in short rocking scans near the top of the peak counting about 60~s/point. To extract the magnetic contribution, these scans were fit using peak profiles of the same width as determined for the underlying nuclear peak above $T_N$. 

Spin wave excitations in \PbV were measured using the Multi-Angle Crystal Spectrometer (MACS) at NIST~\cite{Rodriguez_2008} and the IN12 3-axis instrument at ILL.\cite{SCHMALZL2016}  In both cases the final neutron energy was fixed at E$_f$=3.7~meV. Higher-order beam contamination was suppressed by a cooled BeO filter on MACS and a cooled Be filter on IN12 after the sample. No collimators were used in either of the setups. Both experiments were performed at the base temperature of a dilution refrigerator below 0.1~K. The scattering planes were $(h,k,0)$ and $(0,k,l)$ on MACS and IN12, respectively. The tabulated(MACS) and measured(IN12) energy resolution at the elastic position was correspondingly 0.17~meV and 0.12~meV full width at half maximum. On MACS the data were collected in constant-energy slices at 0.5, 1, 1.5, 1.75, 2 and 2.25~meV energy transfer. Each slice was taken by scanning the scattering angle in the typical range $-108^\circ$ to $83^\circ$ with  $3^\circ$ steps and the sample rotation angle in range $120^\circ$ with  $1^\circ$ steps, while typically  counting $60$~s at each setting.   IN12 data were taken in a series of constant-$q$ scans centered along the $(0,k,0)$, $(0,k,2)$, $(0,2.5,l)$ and $(0,3,l)$ reciprocal-space rods with typical steps of 0.025~meV in energy transfer and counting about 120~s/point.

The high-field ESR  measurements were performed employing  a 16 T   transmission-type  ESR spectrometer,  similar to that  described in Ref. \onlinecite{ZVYAGIN20041}. In our experiments, a set of VDI microwave sources  was used, allowing us to probe magnetic excitations in a very broad, quasi-continuously covered frequency range from approximately 50 to 400 GHz. The experiments were done in the Faraday and Voigt configurations with magnetic field applied along the $a$, $b$, and $c^*$ axes (where the $c^*$ is the direction perpendicular to the $ab$ plane) at a temperature of 1.4 K.

\section{Results}
\subsection{Diffraction}

\begin{figure}
	\includegraphics[width=\columnwidth]{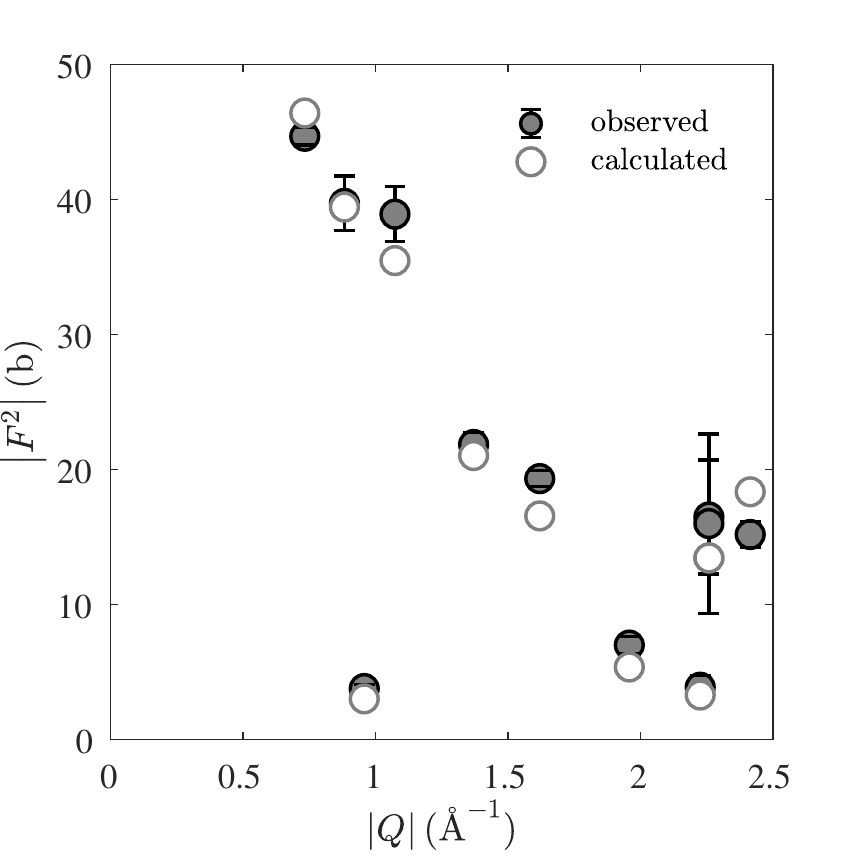}
	\caption{Measured (solid symbols) and calculated (open symbols) squared structure factors of magnetic Bragg reflections in \PbV plotted against momentum transfer. \label{Bragg}}
\end{figure}

The magnetic structure at T = 1.5 K was determined from an analysis of 11 measured magnetic reflections. These were normalized using the scale factor obtained from the least-square fitting of 22 nuclear Bragg peaks measured intensities (R-factor 4.2 \%). A refinement, also using the FULLPROF SUITE package\cite{RODRIGUEZCARVAJAL1993} yielded a unique solution, a collinear CAF-type spin arrangement with moments along the crystallographic $b$ axis [Fig. \ref{crystal}(b)]. The resulting magnetic structure factors are plotted against wave vector in Fig.~\ref{Bragg}, solid symbols. The relative alignment of spins from adjacent V-planes is ferromagnetic. The final R-factor was 5.7\%. The ordered moment was determined to be 0.68(1)~$\mu_B$ per site. The calculated structure factors are plotted in open symbols in   Fig.~\ref{Bragg} for a direct comparison with experiment.

\begin{figure}
\includegraphics[width=\columnwidth]{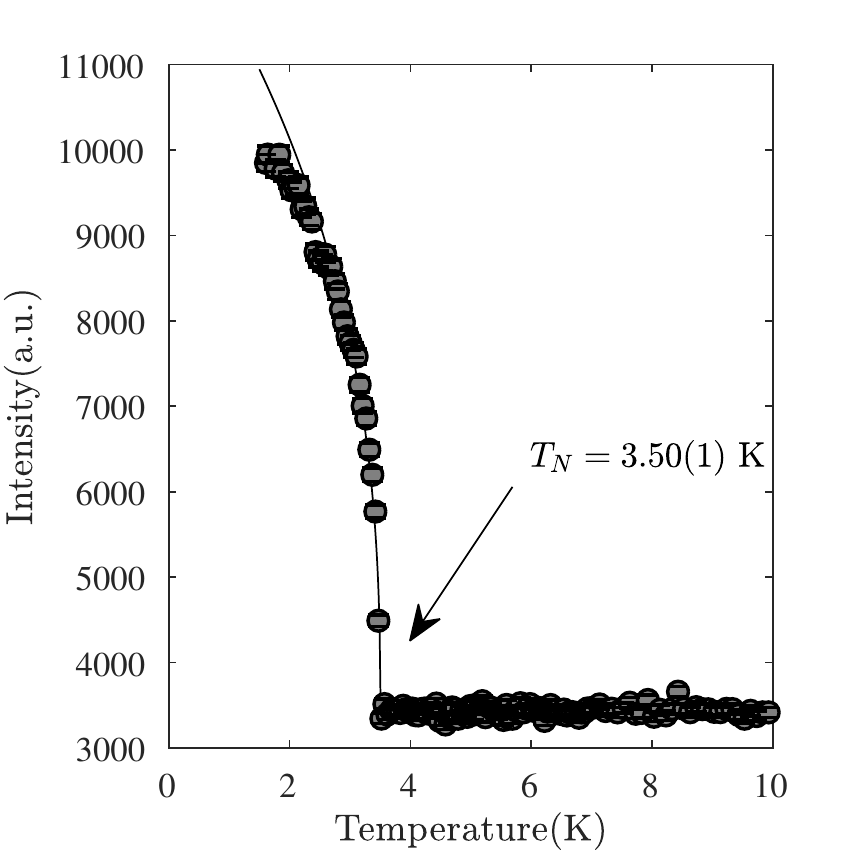}
\caption{Measured temperature dependence of the $(1,0,0)$ magnetic Bragg peak intensity (symbols) and an empirical power law fit to the data as described in the text (solid line).}\label{order}
\end{figure}

In Fig.~\ref{order} we show the temperature dependence of neutron intensity measured at the position of a purely magnetic $(1,0,0)$ Bragg reflection. A simplistic power law fit in the temperature range $2.48<T<9.93$~K yields an ordering temperature of $T_N=3.50(1)$~K and a crude estimate of the order parameter exponent $\beta=0.20(2)$.

\subsection{Inelastic scattering}

\begin{figure*}
	\centering
	\includegraphics[width=\textwidth]{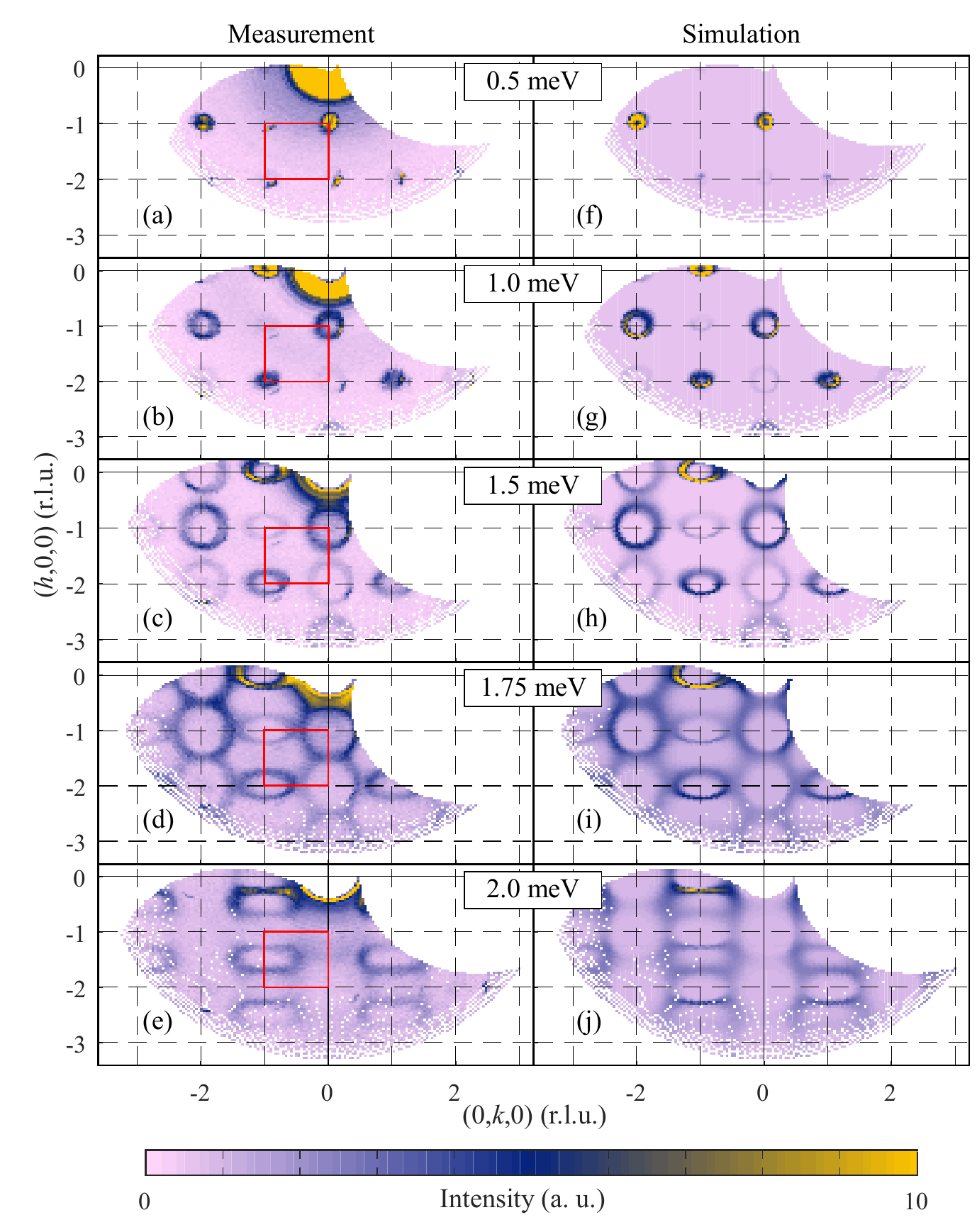}
	\caption{(a)-(e): False color plots of inelastic neutron scattering intensities measured in \PbV at $T<0.1$~K on the MACS spectrometer. Each panel corresponds to a different energy transfer.   (f)-(j): Spin wave theory simulations using fitted parameter values in Table~\ref{distances} and a convolution with the calculated instrument resolution as described in the text. The red rectangle shows the Brillouin zone used in the fit. }
	\label{MACSdata}
\end{figure*}

The first column in Fig.~\ref{MACSdata} shows false color plots of inelastic neutron intensities measured in \PbV at different energy transfers at the MACS instrument. These spectra were modeled using linear spin wave theory, assuming a Heisenberg hamiltonian with five distinct in-plane exchange constants as depicted in Fig.~\ref{crystal}(b). The coupling between V-layers was assumed to be negligible, and all excitation widths were assumed to be resolution limited. The spin wave energies and structure factors were calculated using the program  SpinW.\cite{Toth2015} The neutron polarization factors were based on the collinear magnetic structure described above. The magnetic form factor for V$^{4+}$ was taken in the dipolar approximation as calculated in Ref.~\onlinecite{wilson1992editor}. The thus computed inelastic magnetic neutron scattering cross section was numerically folded with the resolution function of the instrument calculated using the ResLib program.\cite{Reslib} The resolution calculation was done within the Popovici approximation.\cite{Popovici75} The data collected at all energy transfers were fit simultaneously. However, since the process is rather computation intensive, all fits were restricted to a single Brillouin zone shown as a red rectangle at each energy in Fig.~\ref{MACSdata}. The parameters were the five exchange constants, an overall intensity prefactor and a separate constant background at each energy transfer. Treating the exchange constants $J_{1,2}$ and $J_{1,3}$ as independent did not meaningfully improve the quality of the fits. Since they correspond to V-V bond lengths that are practically equal (but not identical by symmetry), in our final analysis they were constrained to be equal. An excellent fit is obtained with parameter values listed in Table~\ref{distances}. This set of Heisenberg exchange constants provides a very good description not only of the data in the target Brillouin zone, but also of  that in the entire experimental range of momentum transfers. Intensities simulated using our fitting model and the final parameter set are shown in false color plots in the right column of Fig.~\ref{MACSdata}.

\begin{table}
	\begin{tabular}{c c c}
		\hline
		\hline
		Bond&V-V-distance(\AA)&$J$(meV)\\
		\hline
		$J_{1,1}$	&	4.42	&		-0.286(2)	\\
		\parbox{0.3\columnwidth}{$J_{1,2}$\\$J_{1,3}$}&\parbox{0.3\columnwidth}{4.66\\4.67}& \parbox{0.3\columnwidth}{-0.389(2)} \\
		$J_{2,1}$	&	6.27	&		1.453(3)	\\
		$J_{2,2}$	&	6.30	&		0.538(2)	\\
		\hline
		\hline
	\end{tabular}
	\caption{Nearest and next nearest neighbor V-V distances in \PbV and exchange parameters obtained from analyzing the inelastic neutron data.}
	\label{distances}
\end{table}

\begin{figure}
\centering
\includegraphics[width=\columnwidth]{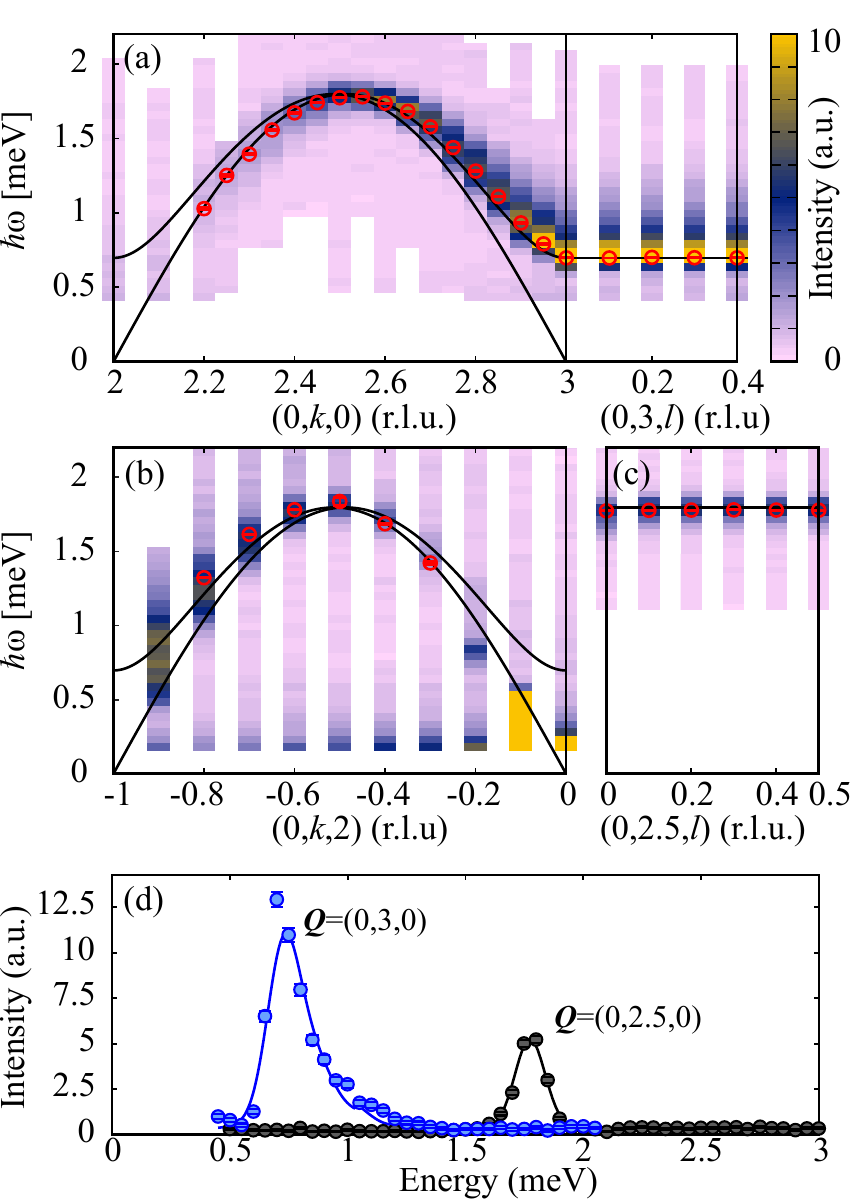}
	\caption{(a)-(c) False color plots of inelastic neutron scattering intensities measured at $T<0.1$~K on IN12.\cite{BettlerIN12data} Red symbols are peak positions in individual scans as obtained through the procedure described in the text. Solid lines show the spin wave dispersion calculated using parameter values in Table~\ref{distances}.
(d) Constant-Q scans at the Brillouin zone center and boundary. Solid lines are fits as described in the text.}	
	\label{IN12data}
\end{figure}

The inelastic intensities collected on IN12 are shown in the false color plots of Fig.~\ref{IN12data}. In their analysis we also employed a combination of SpinW and ResLib, but chose to fit every energy scan separately. To account for resolution (focusing) effects, the spin wave dispersion was calculated using parameters values in Table~\ref{distances}. To independently extract the excitation energy in each scan, we allowed for an additional energy offset relative to this calculated dispersion. A flat background and an intensity scale factor were the two other parameters.  The thus obtained spin wave energies are shown as open circles in Fig.~\ref{IN12data}. The solid lines are a dispersion calculation based on parameters in Table~\ref{distances}. We see an almost perfect agreement. The IN12 data re-affirm the determined values of in-plane exchange constants. In addition, they confirm a total lack of dispersion perpendicular to the planes. From our analysis we can estimate the corresponding bandwidth to be smaller than  20 $\mu$eV.

\subsection{ESR}

\begin{figure}
\includegraphics[width=\columnwidth]{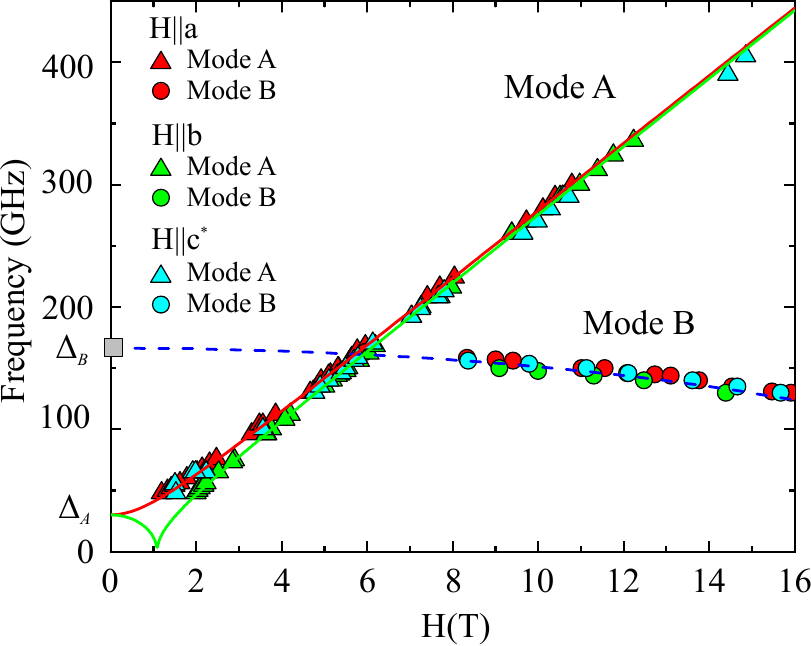}
\caption{\label{fig:ESR}  The frequency-field diagram of ESR excitations measured at $T=1.5$ K for different orientations of magnetic field. Solid lines are the fit results (see text for the details) and the dashed line is a guide to the eye. The zero-field gap as observed through inelastic neutron scattering is shown by a gray square. }
\end{figure}

High-field ESR experiments revealed two ESR modes, whose frequency-field diagrams are shown in Fig.~\ref{fig:ESR}. The frequency-field dependence of  mode A can be fit 
using  the equation $\hbar\omega_A = 
\sqrt{|\Delta_A^2 \pm (g\mu_B H)^2}|$
, where $\hbar$ is the Planck constant, $\omega_A$ is the excitation frequency, $\mu_B$ is the Bohr magneton, and  $g=1.98$ is the $g$ factor.  The fit results are shown in Fig.~\ref{fig:ESR}  by the red and green solid lines, indicating a zero-field gap $\Delta_A=30$ GHz ($\sim 0.12$ meV), which is present in  Pb$_2$VO(PO$_4$)$_2$  presumably   due to a finite in-plane anisotropy. Mode  B  can be identified as an exchange mode with a gap $\Delta_B=168 $ GHz ($\sim 0.69$ meV), extrapolated from the frequency-field dependence. The gap size is in excellent agreement with results of inelastic neutron scattering (Fig. 5).

\section{Discussion}
While our results confirm the two-dimensional nature and a CAF-type magnetic structure for \PbV, they also unambiguously show that a square lattice description with only two exchange parameters is insufficient to describe the magnetism in this compound. The nearest-neighbor coupling constants are indeed all ferromagnetic, but alternate substantially along the $a$ axis. The most significant consequence is the splitting of the spin wave spectrum into two separate branches as seen in Fig.~\ref{IN12data}. This produces an additional gapped excitation at each zone-center, as also seen with ESR.\cite{Schmidt2010} Next-nearest-neighbor interactions are AF, but differ by almost a factor of three along the two diagonals. 

Our data also show that the material is less frustrated than originally thought. A good measure of frustration is the ratio $\alpha^{-1}=1+S^2\sum |J|/E_{\mathrm{cl}}$, where $\sum |J|$ is the sum of absolute values all exchange constants and $E_{\mathrm{cl}}$ is the classical ground state energy. In all cases $\alpha^{-1}\le 0$ with $\alpha=0$ corresponding to an absence of frustration.
For the ferro-antiferromagnet square lattice CAF, $\alpha^{-1}={J_1}/{J_2}$. In our exchange model for \PbV,  $E_{\mathrm{cl}}/S^2=J_{1,3}-J_{1,1}/2-J_{1,2}/2-J_{2,1}-J_{2,2}$ and $\alpha^{-1}\approx -0.32(1)$.  Previous estimates based on an assumed square lattice model\cite{Tsirlin2009b} correspond to a stronger frustration $\alpha^{-1}\approx -0.62$. A weak frustration in \PbV is consistent with a rather large observed ordered moment. The latter is similar to that found in strongly 2-dimensional but unfrustrated $S=1/2$ AFs such as the cuprates.\cite{TRANQUADA1989}

\section{Conclusion}
While  \PbV and probably all related layered vanadophosphates are indeed highly 2-dimensional and feature competing ferromagnetic and antiferromagnetic interactions   the actual spin Hamiltonian in these systems may be more sensitive to structural details than originally thought. 

\acknowledgments
Access to MACS was provided by the Center for High Resolution Neutron Scattering, a partnership between the National Institute of Standards and Technology and the National Science Foundation under Agreement No. DMR-1508249. This work was supported by the Deutsche Forschungsgemeinschaft through the
projects ZV6/2-2, SFB 1143, and the W\"{u}rzburg-Dresden Cluster of Excellence
on Complexity and Topology in Quantum Matter - $ct.qmat$ (EXC 2147, project-id
39085490).  We acknowledge support by  the HLD at HZDR, member of the
European Magnetic Field Laboratory (EMFL). This work was also supported by the Swiss State Secretariat for Education, Research and Innovation (SERI) through a CRG-grant. This work is partially supported by the Swiss National Science Foundation under Division II.
The authors would like to thank  K.~Povarov  for fruitful discussions.
\bibliography{SB_bib}

\begin{thebibliography}{27}%
\makeatletter
\providecommand \@ifxundefined [1]{%
 \@ifx{#1\undefined}
}%
\providecommand \@ifnum [1]{%
 \ifnum #1\expandafter \@firstoftwo
 \else \expandafter \@secondoftwo
 \fi
}%
\providecommand \@ifx [1]{%
 \ifx #1\expandafter \@firstoftwo
 \else \expandafter \@secondoftwo
 \fi
}%
\providecommand \natexlab [1]{#1}%
\providecommand \enquote  [1]{``#1''}%
\providecommand \bibnamefont  [1]{#1}%
\providecommand \bibfnamefont [1]{#1}%
\providecommand \citenamefont [1]{#1}%
\providecommand \href@noop [0]{\@secondoftwo}%
\providecommand \href [0]{\begingroup \@sanitize@url \@href}%
\providecommand \@href[1]{\@@startlink{#1}\@@href}%
\providecommand \@@href[1]{\endgroup#1\@@endlink}%
\providecommand \@sanitize@url [0]{\catcode `\\12\catcode `\$12\catcode
  `\&12\catcode `\#12\catcode `\^12\catcode `\_12\catcode `\%12\relax}%
\providecommand \@@startlink[1]{}%
\providecommand \@@endlink[0]{}%
\providecommand \url  [0]{\begingroup\@sanitize@url \@url }%
\providecommand \@url [1]{\endgroup\@href {#1}{\urlprefix }}%
\providecommand \urlprefix  [0]{URL }%
\providecommand \Eprint [0]{\href }%
\providecommand \doibase [0]{http://dx.doi.org/}%
\providecommand \selectlanguage [0]{\@gobble}%
\providecommand \bibinfo  [0]{\@secondoftwo}%
\providecommand \bibfield  [0]{\@secondoftwo}%
\providecommand \translation [1]{[#1]}%
\providecommand \BibitemOpen [0]{}%
\providecommand \bibitemStop [0]{}%
\providecommand \bibitemNoStop [0]{.\EOS\space}%
\providecommand \EOS [0]{\spacefactor3000\relax}%
\providecommand \BibitemShut  [1]{\csname bibitem#1\endcsname}%
\let\auto@bib@innerbib\@empty
%</preamble>
\bibitem [{\citenamefont {Starykh}(2015)}]{Starykh2015}%
  \BibitemOpen
  \bibfield  {author} {\bibinfo {author} {\bibfnamefont {O.~A.}\ \bibnamefont
  {Starykh}},\ }\href {\doibase 10.1088/0034-4885/78/5/052502} {\bibfield
  {journal} {\bibinfo  {journal} {Reports on Progress in Physics}\ }\textbf
  {\bibinfo {volume} {78}},\ \bibinfo {pages} {052502} (\bibinfo {year}
  {2015})}\BibitemShut {NoStop}%
\bibitem [{\citenamefont {Shannon}\ \emph {et~al.}(2006)\citenamefont
  {Shannon}, \citenamefont {Momoi},\ and\ \citenamefont
  {Sindzingre}}]{Shannon2006}%
  \BibitemOpen
  \bibfield  {author} {\bibinfo {author} {\bibfnamefont {N.}~\bibnamefont
  {Shannon}}, \bibinfo {author} {\bibfnamefont {T.}~\bibnamefont {Momoi}}, \
  and\ \bibinfo {author} {\bibfnamefont {P.}~\bibnamefont {Sindzingre}},\
  }\href {\doibase 10.1103/PhysRevLett.96.027213} {\bibfield  {journal}
  {\bibinfo  {journal} {Phys. Rev. Lett.}\ }\textbf {\bibinfo {volume} {96}},\
  \bibinfo {pages} {027213} (\bibinfo {year} {2006})}\BibitemShut {NoStop}%
\bibitem [{\citenamefont {Ueda}\ and\ \citenamefont
  {Totsuka}(2007)}]{Ueda2007}%
  \BibitemOpen
  \bibfield  {author} {\bibinfo {author} {\bibfnamefont {H.~T.}\ \bibnamefont
  {Ueda}}\ and\ \bibinfo {author} {\bibfnamefont {K.}~\bibnamefont {Totsuka}},\
  }\href {\doibase 10.1103/PhysRevB.76.214428} {\bibfield  {journal} {\bibinfo
  {journal} {Phys. Rev. B}\ }\textbf {\bibinfo {volume} {76}},\ \bibinfo
  {pages} {214428} (\bibinfo {year} {2007})}\BibitemShut {NoStop}%
\bibitem [{\citenamefont {Shindou}\ \emph {et~al.}(2013)\citenamefont
  {Shindou}, \citenamefont {Yunoki},\ and\ \citenamefont
  {Momoi}}]{Shindou2013}%
  \BibitemOpen
  \bibfield  {author} {\bibinfo {author} {\bibfnamefont {R.}~\bibnamefont
  {Shindou}}, \bibinfo {author} {\bibfnamefont {S.}~\bibnamefont {Yunoki}}, \
  and\ \bibinfo {author} {\bibfnamefont {T.}~\bibnamefont {Momoi}},\ }\href
  {\doibase 10.1103/PhysRevB.87.054429} {\bibfield  {journal} {\bibinfo
  {journal} {Phys. Rev. B}\ }\textbf {\bibinfo {volume} {87}},\ \bibinfo
  {pages} {054429} (\bibinfo {year} {2013})}\BibitemShut {NoStop}%
\bibitem [{\citenamefont {Ueda}\ and\ \citenamefont {Momoi}(2013)}]{Ueda2013}%
  \BibitemOpen
  \bibfield  {author} {\bibinfo {author} {\bibfnamefont {H.~T.}\ \bibnamefont
  {Ueda}}\ and\ \bibinfo {author} {\bibfnamefont {T.}~\bibnamefont {Momoi}},\
  }\href {\doibase 10.1103/PhysRevB.87.144417} {\bibfield  {journal} {\bibinfo
  {journal} {Phys. Rev. B}\ }\textbf {\bibinfo {volume} {87}},\ \bibinfo
  {pages} {144417} (\bibinfo {year} {2013})}\BibitemShut {NoStop}%
\bibitem [{\citenamefont {Ueda}(2015)}]{Ueda2015}%
  \BibitemOpen
  \bibfield  {author} {\bibinfo {author} {\bibfnamefont {H.~T.}\ \bibnamefont
  {Ueda}},\ }\href {\doibase 10.7566/JPSJ.84.023601} {\bibfield  {journal}
  {\bibinfo  {journal} {Journal of the Physical Society of Japan}\ }\textbf
  {\bibinfo {volume} {84}},\ \bibinfo {pages} {023601} (\bibinfo {year}
  {2015})}\BibitemShut {NoStop}%
\bibitem [{\citenamefont {Tsirlin}\ and\ \citenamefont
  {Rosner}(2009)}]{Tsirlin2009}%
  \BibitemOpen
  \bibfield  {author} {\bibinfo {author} {\bibfnamefont {A.~A.}\ \bibnamefont
  {Tsirlin}}\ and\ \bibinfo {author} {\bibfnamefont {H.}~\bibnamefont
  {Rosner}},\ }\href {\doibase 10.1103/PhysRevB.79.214417} {\bibfield
  {journal} {\bibinfo  {journal} {Phys. Rev. B}\ }\textbf {\bibinfo {volume}
  {79}},\ \bibinfo {pages} {214417} (\bibinfo {year} {2009})}\BibitemShut
  {NoStop}%
\bibitem [{\citenamefont {Skoulatos}\ \emph {et~al.}(2007)\citenamefont
  {Skoulatos}, \citenamefont {Goff}, \citenamefont {Shannon}, \citenamefont
  {Kaul}, \citenamefont {Geibel}, \citenamefont {Murani}, \citenamefont
  {Enderle},\ and\ \citenamefont {Wildes}}]{Skoulatos2007}%
  \BibitemOpen
  \bibfield  {author} {\bibinfo {author} {\bibfnamefont {M.}~\bibnamefont
  {Skoulatos}}, \bibinfo {author} {\bibfnamefont {J.~P.}\ \bibnamefont {Goff}},
  \bibinfo {author} {\bibfnamefont {N.}~\bibnamefont {Shannon}}, \bibinfo
  {author} {\bibfnamefont {E.~E.}\ \bibnamefont {Kaul}}, \bibinfo {author}
  {\bibfnamefont {C.}~\bibnamefont {Geibel}}, \bibinfo {author} {\bibfnamefont
  {A.~P.}\ \bibnamefont {Murani}}, \bibinfo {author} {\bibfnamefont
  {M.}~\bibnamefont {Enderle}}, \ and\ \bibinfo {author} {\bibfnamefont
  {A.~R.}\ \bibnamefont {Wildes}},\ }\href@noop {} {\bibfield  {journal}
  {\bibinfo  {journal} {Journal of Magnetism and Magnetic Materials}\ }\textbf
  {\bibinfo {volume} {310}},\ \bibinfo {pages} {1257} (\bibinfo {year}
  {2007})}\BibitemShut {NoStop}%
\bibitem [{\citenamefont {Skoulatos}\ \emph {et~al.}(2009)\citenamefont
  {Skoulatos}, \citenamefont {Goff}, \citenamefont {Geibel}, \citenamefont
  {Kaul}, \citenamefont {Nath}, \citenamefont {Shannon}, \citenamefont
  {Schmidt}, \citenamefont {Murani}, \citenamefont {Deen}, \citenamefont
  {Enderle},\ and\ \citenamefont {Wildes}}]{Skoulatos2009}%
  \BibitemOpen
  \bibfield  {author} {\bibinfo {author} {\bibfnamefont {M.}~\bibnamefont
  {Skoulatos}}, \bibinfo {author} {\bibfnamefont {J.~P.}\ \bibnamefont {Goff}},
  \bibinfo {author} {\bibfnamefont {C.}~\bibnamefont {Geibel}}, \bibinfo
  {author} {\bibfnamefont {E.~E.}\ \bibnamefont {Kaul}}, \bibinfo {author}
  {\bibfnamefont {R.}~\bibnamefont {Nath}}, \bibinfo {author} {\bibfnamefont
  {N.}~\bibnamefont {Shannon}}, \bibinfo {author} {\bibfnamefont
  {B.}~\bibnamefont {Schmidt}}, \bibinfo {author} {\bibfnamefont {A.~P.}\
  \bibnamefont {Murani}}, \bibinfo {author} {\bibfnamefont {P.~P.}\
  \bibnamefont {Deen}}, \bibinfo {author} {\bibfnamefont {M.}~\bibnamefont
  {Enderle}}, \ and\ \bibinfo {author} {\bibfnamefont {A.~R.}\ \bibnamefont
  {Wildes}},\ }\href {\doibase 10.1209/0295-5075/88/57005} {\bibfield
  {journal} {\bibinfo  {journal} {Europys. Lett.}\ }\textbf {\bibinfo {volume}
  {88}},\ \bibinfo {pages} {57005} (\bibinfo {year} {2009})}\BibitemShut
  {NoStop}%
\bibitem [{\citenamefont {Kaul}\ \emph {et~al.}(2004)\citenamefont {Kaul},
  \citenamefont {Rosner}, \citenamefont {Shannon}, \citenamefont
  {Shpanchenko},\ and\ \citenamefont {Geibel}}]{Kaul2004}%
  \BibitemOpen
  \bibfield  {author} {\bibinfo {author} {\bibfnamefont {E.~E.}\ \bibnamefont
  {Kaul}}, \bibinfo {author} {\bibfnamefont {H.}~\bibnamefont {Rosner}},
  \bibinfo {author} {\bibfnamefont {N.}~\bibnamefont {Shannon}}, \bibinfo
  {author} {\bibfnamefont {R.~V.}\ \bibnamefont {Shpanchenko}}, \ and\ \bibinfo
  {author} {\bibfnamefont {C.}~\bibnamefont {Geibel}},\ }\href {\doibase
  10.1016/j.jmmm.2003.12.002} {\bibfield  {journal} {\bibinfo  {journal}
  {Journal of Magnetism and Magnetic Materials}\ } (\bibinfo {year} {2004}),\
  10.1016/j.jmmm.2003.12.002}\BibitemShut {NoStop}%
\bibitem [{\citenamefont {F\"orster}\ \emph {et~al.}(2013)\citenamefont
  {F\"orster}, \citenamefont {Garcia}, \citenamefont {Gruner}, \citenamefont
  {Kaul}, \citenamefont {Schmidt}, \citenamefont {Geibel},\ and\ \citenamefont
  {Sichelschmidt}}]{Forster2013}%
  \BibitemOpen
  \bibfield  {author} {\bibinfo {author} {\bibfnamefont {T.}~\bibnamefont
  {F\"orster}}, \bibinfo {author} {\bibfnamefont {F.~A.}\ \bibnamefont
  {Garcia}}, \bibinfo {author} {\bibfnamefont {T.}~\bibnamefont {Gruner}},
  \bibinfo {author} {\bibfnamefont {E.~E.}\ \bibnamefont {Kaul}}, \bibinfo
  {author} {\bibfnamefont {B.}~\bibnamefont {Schmidt}}, \bibinfo {author}
  {\bibfnamefont {C.}~\bibnamefont {Geibel}}, \ and\ \bibinfo {author}
  {\bibfnamefont {J.}~\bibnamefont {Sichelschmidt}},\ }\href {\doibase
  10.1103/PhysRevB.87.180401} {\bibfield  {journal} {\bibinfo  {journal} {Phys.
  Rev. B}\ }\textbf {\bibinfo {volume} {87}},\ \bibinfo {pages} {180401(R)}
  (\bibinfo {year} {2013})}\BibitemShut {NoStop}%
\bibitem [{\citenamefont {Nath}\ \emph {et~al.}(2009)\citenamefont {Nath},
  \citenamefont {Furukawa}, \citenamefont {Borsa}, \citenamefont {Kaul},
  \citenamefont {Baenitz}, \citenamefont {Geibel},\ and\ \citenamefont
  {Johnston}}]{Nath2009}%
  \BibitemOpen
  \bibfield  {author} {\bibinfo {author} {\bibfnamefont {R.}~\bibnamefont
  {Nath}}, \bibinfo {author} {\bibfnamefont {Y.}~\bibnamefont {Furukawa}},
  \bibinfo {author} {\bibfnamefont {F.}~\bibnamefont {Borsa}}, \bibinfo
  {author} {\bibfnamefont {E.~E.}\ \bibnamefont {Kaul}}, \bibinfo {author}
  {\bibfnamefont {M.}~\bibnamefont {Baenitz}}, \bibinfo {author} {\bibfnamefont
  {C.}~\bibnamefont {Geibel}}, \ and\ \bibinfo {author} {\bibfnamefont {D.~C.}\
  \bibnamefont {Johnston}},\ }\href {\doibase 10.1103/PhysRevB.80.214430}
  {\bibfield  {journal} {\bibinfo  {journal} {Phys. Rev. B}\ }\textbf {\bibinfo
  {volume} {80}},\ \bibinfo {pages} {214430} (\bibinfo {year}
  {2009})}\BibitemShut {NoStop}%
\bibitem [{\citenamefont {Kaul}(2005)}]{Kaul2005}%
  \BibitemOpen
  \bibfield  {author} {\bibinfo {author} {\bibfnamefont {E.~E.}\ \bibnamefont
  {Kaul}},\ }\emph {\bibinfo {title} {Experimental Investigation of New
  Low-Dimensional Spin Systems in Vanadium Oxides}},\ \href@noop {} {Ph.D.
  thesis},\ \bibinfo  {school} {Technische Universit\"at Dresden} (\bibinfo
  {year} {2005})\BibitemShut {NoStop}%
\bibitem [{\citenamefont {Shpanchenko}\ \emph {et~al.}(2006)\citenamefont
  {Shpanchenko}, \citenamefont {Kaul}, \citenamefont {Geibel},\ and\
  \citenamefont {Antipov}}]{Shpanchenko2006}%
  \BibitemOpen
  \bibfield  {author} {\bibinfo {author} {\bibfnamefont {V.}~\bibnamefont
  {Shpanchenko}}, \bibinfo {author} {\bibfnamefont {E.~E.}\ \bibnamefont
  {Kaul}}, \bibinfo {author} {\bibfnamefont {C.}~\bibnamefont {Geibel}}, \ and\
  \bibinfo {author} {\bibfnamefont {E.~V.}\ \bibnamefont {Antipov}},\
  }\href@noop {} {\bibfield  {journal} {\bibinfo  {journal} {Acta Crystallog.
  C}\ }\textbf {\bibinfo {volume} {62}},\ \bibinfo {pages} {88} (\bibinfo
  {year} {2006})}\BibitemShut {NoStop}%
\bibitem [{Note1()}]{Note1}%
  \BibitemOpen
  \bibinfo {note} {The identification of any commercial product or trade name
  does not imply endorsement or recommendation by the National Institute of
  Standards and Technology.}\BibitemShut {Stop}%
\bibitem [{\citenamefont {Rodriguez}\ \emph {et~al.}(2008)\citenamefont
  {Rodriguez}, \citenamefont {Adler}, \citenamefont {Brand}, \citenamefont
  {Broholm}, \citenamefont {Cook}, \citenamefont {Brocker}, \citenamefont
  {Hammond}, \citenamefont {Huang}, \citenamefont {Hundertmark}, \citenamefont
  {Lynn}, \citenamefont {Maliszewskyj}, \citenamefont {Moyer}, \citenamefont
  {Orndorff}, \citenamefont {Pierce}, \citenamefont {Pike}, \citenamefont
  {Scharfstein}, \citenamefont {Smee},\ and\ \citenamefont
  {Vilaseca}}]{Rodriguez_2008}%
  \BibitemOpen
  \bibfield  {author} {\bibinfo {author} {\bibfnamefont {J.~A.}\ \bibnamefont
  {Rodriguez}}, \bibinfo {author} {\bibfnamefont {D.~M.}\ \bibnamefont
  {Adler}}, \bibinfo {author} {\bibfnamefont {P.~C.}\ \bibnamefont {Brand}},
  \bibinfo {author} {\bibfnamefont {C.}~\bibnamefont {Broholm}}, \bibinfo
  {author} {\bibfnamefont {J.~C.}\ \bibnamefont {Cook}}, \bibinfo {author}
  {\bibfnamefont {C.}~\bibnamefont {Brocker}}, \bibinfo {author} {\bibfnamefont
  {R.}~\bibnamefont {Hammond}}, \bibinfo {author} {\bibfnamefont
  {Z.}~\bibnamefont {Huang}}, \bibinfo {author} {\bibfnamefont
  {P.}~\bibnamefont {Hundertmark}}, \bibinfo {author} {\bibfnamefont {J.~W.}\
  \bibnamefont {Lynn}}, \bibinfo {author} {\bibfnamefont {N.~C.}\ \bibnamefont
  {Maliszewskyj}}, \bibinfo {author} {\bibfnamefont {J.}~\bibnamefont {Moyer}},
  \bibinfo {author} {\bibfnamefont {J.}~\bibnamefont {Orndorff}}, \bibinfo
  {author} {\bibfnamefont {D.}~\bibnamefont {Pierce}}, \bibinfo {author}
  {\bibfnamefont {T.~D.}\ \bibnamefont {Pike}}, \bibinfo {author}
  {\bibfnamefont {G.}~\bibnamefont {Scharfstein}}, \bibinfo {author}
  {\bibfnamefont {S.~A.}\ \bibnamefont {Smee}}, \ and\ \bibinfo {author}
  {\bibfnamefont {R.}~\bibnamefont {Vilaseca}},\ }\href {\doibase
  10.1088/0957-0233/19/3/034023} {\bibfield  {journal} {\bibinfo  {journal}
  {Measurement Science and Technology}\ }\textbf {\bibinfo {volume} {19}},\
  \bibinfo {pages} {034023} (\bibinfo {year} {2008})}\BibitemShut {NoStop}%
\bibitem [{\citenamefont {Schmalzl}\ \emph {et~al.}(2016)\citenamefont
  {Schmalzl}, \citenamefont {Schmidt}, \citenamefont {Raymond}, \citenamefont
  {Feilbach}, \citenamefont {Mounier}, \citenamefont {Vettard},\ and\
  \citenamefont {Brückel}}]{SCHMALZL2016}%
  \BibitemOpen
  \bibfield  {author} {\bibinfo {author} {\bibfnamefont {K.}~\bibnamefont
  {Schmalzl}}, \bibinfo {author} {\bibfnamefont {W.}~\bibnamefont {Schmidt}},
  \bibinfo {author} {\bibfnamefont {S.}~\bibnamefont {Raymond}}, \bibinfo
  {author} {\bibfnamefont {H.}~\bibnamefont {Feilbach}}, \bibinfo {author}
  {\bibfnamefont {C.}~\bibnamefont {Mounier}}, \bibinfo {author} {\bibfnamefont
  {B.}~\bibnamefont {Vettard}}, \ and\ \bibinfo {author} {\bibfnamefont
  {T.}~\bibnamefont {Brückel}},\ }\href {\doibase
  https://doi.org/10.1016/j.nima.2016.02.067} {\bibfield  {journal} {\bibinfo
  {journal} {Nuclear Instruments and Methods in Physics Research Section A:
  Accelerators, Spectrometers, Detectors and Associated Equipment}\ }\textbf
  {\bibinfo {volume} {819}},\ \bibinfo {pages} {89 } (\bibinfo {year}
  {2016})}\BibitemShut {NoStop}%
\bibitem [{\citenamefont {Zvyagin}\ \emph {et~al.}(2004)\citenamefont
  {Zvyagin}, \citenamefont {Krzystek}, \citenamefont {van Loosdrecht},
  \citenamefont {Dhalenne},\ and\ \citenamefont {Revcolevschi}}]{ZVYAGIN20041}%
  \BibitemOpen
  \bibfield  {author} {\bibinfo {author} {\bibfnamefont {S.}~\bibnamefont
  {Zvyagin}}, \bibinfo {author} {\bibfnamefont {J.}~\bibnamefont {Krzystek}},
  \bibinfo {author} {\bibfnamefont {P.}~\bibnamefont {van Loosdrecht}},
  \bibinfo {author} {\bibfnamefont {G.}~\bibnamefont {Dhalenne}}, \ and\
  \bibinfo {author} {\bibfnamefont {A.}~\bibnamefont {Revcolevschi}},\ }\href
  {\doibase https://doi.org/10.1016/j.physb.2004.01.009} {\bibfield  {journal}
  {\bibinfo  {journal} {Physica B: Condensed Matter}\ }\textbf {\bibinfo
  {volume} {346-347}},\ \bibinfo {pages} {1 } (\bibinfo {year} {2004})},\
  \bibinfo {note} {proceedings of the 7th International Symposium on Research
  in High Magnetic Fields}\BibitemShut {NoStop}%
\bibitem [{\citenamefont {Rodríguez-Carvajal}(1993)}]{RODRIGUEZCARVAJAL1993}%
  \BibitemOpen
  \bibfield  {author} {\bibinfo {author} {\bibfnamefont {J.}~\bibnamefont
  {Rodríguez-Carvajal}},\ }\href {\doibase
  https://doi.org/10.1016/0921-4526(93)90108-I} {\bibfield  {journal} {\bibinfo
   {journal} {Physica B: Condensed Matter}\ }\textbf {\bibinfo {volume}
  {192}},\ \bibinfo {pages} {55 } (\bibinfo {year} {1993})}\BibitemShut
  {NoStop}%
\bibitem [{\citenamefont {Toth}\ and\ \citenamefont {Lake}(2015)}]{Toth2015}%
  \BibitemOpen
  \bibfield  {author} {\bibinfo {author} {\bibfnamefont {S.}~\bibnamefont
  {Toth}}\ and\ \bibinfo {author} {\bibfnamefont {B.}~\bibnamefont {Lake}},\
  }\href {http://stacks.iop.org/0953-8984/27/i=16/a=166002} {\bibfield
  {journal} {\bibinfo  {journal} {Journal of Physics: Condensed Matter}\
  }\textbf {\bibinfo {volume} {27}},\ \bibinfo {pages} {166002} (\bibinfo
  {year} {2015})}\BibitemShut {NoStop}%
\bibitem [{\citenamefont {Wilson}\ and\ \citenamefont
  {Prince}(1992)}]{wilson1992editor}%
  \BibitemOpen
  \bibinfo {editor} {\bibfnamefont {A.}~\bibnamefont {Wilson}}\ and\ \bibinfo
  {editor} {\bibfnamefont {E.}~\bibnamefont {Prince}},\ eds.,\ \href@noop {}
  {\emph {\bibinfo {title} {International Tables for Crystallography, Vol.
  C}}}\ (\bibinfo  {publisher} {Kluwer Academic Publishers},\ \bibinfo
  {address} {Dordrecht, Boston, London},\ \bibinfo {year} {1992})\BibitemShut
  {NoStop}%
\bibitem [{\citenamefont {Zheludev}(2009)}]{Reslib}%
  \BibitemOpen
  \bibfield  {author} {\bibinfo {author} {\bibfnamefont {A.}~\bibnamefont
  {Zheludev}},\ }\href@noop {} {\enquote {\bibinfo {title} {Reslib resolution
  library for matlab},}\ }\bibinfo {howpublished}
  {http://www.neutron.ethz.ch/research/resources/reslib} (\bibinfo {year}
  {2009})\BibitemShut {NoStop}%
\bibitem [{\citenamefont {Popovici}(1975)}]{Popovici75}%
  \BibitemOpen
  \bibfield  {author} {\bibinfo {author} {\bibfnamefont {M.}~\bibnamefont
  {Popovici}},\ }\href@noop {} {\bibfield  {journal} {\bibinfo  {journal} {Acta
  Cryst.}\ }\textbf {\bibinfo {volume} {A31}},\ \bibinfo {pages} {507}
  (\bibinfo {year} {1975})}\BibitemShut {NoStop}%
\bibitem [{\citenamefont {Bettler}\ \emph {et~al.}(2018)\citenamefont
  {Bettler}, \citenamefont {Feng}, \citenamefont {Gvasaliya}, \citenamefont
  {Landolt}, \citenamefont {Raymond},\ and\ \citenamefont
  {Zheludev}}]{BettlerIN12data}%
  \BibitemOpen
  \bibfield  {author} {\bibinfo {author} {\bibfnamefont {S.}~\bibnamefont
  {Bettler}}, \bibinfo {author} {\bibfnamefont {Y.}~\bibnamefont {Feng}},
  \bibinfo {author} {\bibfnamefont {S.}~\bibnamefont {Gvasaliya}}, \bibinfo
  {author} {\bibfnamefont {F.}~\bibnamefont {Landolt}}, \bibinfo {author}
  {\bibfnamefont {S.}~\bibnamefont {Raymond}}, \ and\ \bibinfo {author}
  {\bibfnamefont {A.}~\bibnamefont {Zheludev}},\ }\href@noop {} {\enquote
  {\bibinfo {title} {Spin excitations in {Pb$_2$VO(PO$_4$)$_2$} single
  crystals},}\ }\bibinfo {howpublished} {Institut Laue-Langevin (ILL)}
  (\bibinfo {year} {2018}),\ \bibinfo {note}
  {10.5291/ILL-DATA.CRG-2513}\BibitemShut {NoStop}%
\bibitem [{\citenamefont {Schmidt}\ \emph {et~al.}(2010)\citenamefont
  {Schmidt}, \citenamefont {Siahatgar},\ and\ \citenamefont
  {Thalmeier}}]{Schmidt2010}%
  \BibitemOpen
  \bibfield  {author} {\bibinfo {author} {\bibfnamefont {B.}~\bibnamefont
  {Schmidt}}, \bibinfo {author} {\bibfnamefont {M.}~\bibnamefont {Siahatgar}},
  \ and\ \bibinfo {author} {\bibfnamefont {P.}~\bibnamefont {Thalmeier}},\
  }\href {\doibase 10.1103/PhysRevB.81.165101} {\bibfield  {journal} {\bibinfo
  {journal} {Phys. Rev. B}\ }\textbf {\bibinfo {volume} {81}},\ \bibinfo
  {pages} {165101} (\bibinfo {year} {2010})}\BibitemShut {NoStop}%
\bibitem [{\citenamefont {Tsirlin}\ \emph {et~al.}(2009)\citenamefont
  {Tsirlin}, \citenamefont {Schmidt}, \citenamefont {Skourski}, \citenamefont
  {Nath}, \citenamefont {Geibel},\ and\ \citenamefont {Rosner}}]{Tsirlin2009b}%
  \BibitemOpen
  \bibfield  {author} {\bibinfo {author} {\bibfnamefont {A.~A.}\ \bibnamefont
  {Tsirlin}}, \bibinfo {author} {\bibfnamefont {B.}~\bibnamefont {Schmidt}},
  \bibinfo {author} {\bibfnamefont {Y.}~\bibnamefont {Skourski}}, \bibinfo
  {author} {\bibfnamefont {R.}~\bibnamefont {Nath}}, \bibinfo {author}
  {\bibfnamefont {C.}~\bibnamefont {Geibel}}, \ and\ \bibinfo {author}
  {\bibfnamefont {H.}~\bibnamefont {Rosner}},\ }\href {\doibase
  10.1103/PhysRevB.80.132407} {\bibfield  {journal} {\bibinfo  {journal} {Phys.
  Rev. B}\ }\textbf {\bibinfo {volume} {80}},\ \bibinfo {pages} {132407}
  (\bibinfo {year} {2009})}\BibitemShut {NoStop}%
\bibitem [{\citenamefont {Tranquada}\ and\ \citenamefont
  {Shirane}(1989)}]{TRANQUADA1989}%
  \BibitemOpen
  \bibfield  {author} {\bibinfo {author} {\bibfnamefont {J.}~\bibnamefont
  {Tranquada}}\ and\ \bibinfo {author} {\bibfnamefont {G.}~\bibnamefont
  {Shirane}},\ }\href {\doibase https://doi.org/10.1016/0921-4534(89)90491-7}
  {\bibfield  {journal} {\bibinfo  {journal} {Physica C: Superconductivity and
  its Applications}\ }\textbf {\bibinfo {volume} {162-164}},\ \bibinfo {pages}
  {849 } (\bibinfo {year} {1989})}\BibitemShut {NoStop}%
\end{thebibliography}%

\end{document}